\documentclass[prd,twocolumn,groupedaddress]{revtex4-1}

\usepackage{graphicx}
\usepackage{xcolor}
\newcommand{\be}{\begin{eqnarray}}
\newcommand{\ee}{\end{eqnarray}}

\begin{document}

\title{Constraining Non-Cold Dark Matter Models with the Global 21-cm Signal}

\author{Aurel Schneider}
\email{aurel.schneider@phys.ethz.ch}
\affiliation{Institute for Particle Physics and Astrophysics, Department of Physics, ETH Zurich, Wolfgang-Pauli-Strasse 27, 8093, Zurich, Switzerland}

\begin{abstract}
Any particle dark matter (DM) scenario featuring a suppressed power spectrum of astrophysical relevance results in a delay of galaxy formation. As a consequence, such scenarios can be constrained using the global 21-cm absorption signal initiated by the UV radiation of the first stars. The Experiment to Detect the Global Epoch of Reionization Signature (EDGES) recently reported the first detection of such an absorption signal at redshift $\sim 17$. While its amplitude might indicate the need for new physics, we solely focus on the timing of the signal to test non-cold DM models. Assuming conservative limits for the stellar-to-baryon fraction ($f_{*}<0.03$) and for the minimum cooling temperature ($T_{\rm vir}>10^3$ Kelvin) motivated by radiation-hydrodynamic simulations, we are able to derive unprecedented constraints on a variety of non-cold DM models. For example, the mass of thermal warm DM is limited to $m_{\rm TH}>6.1$ keV, while mixed DM scenarios (featuring a cold and a hot component) are constrained to a hot DM fraction below 17 percent. The ultra-light axion DM model is limited to masses $m_{a}>8\times10^{-21}$ eV, a regime where its wave-like nature is pushed far below the kiloparsec scale. Finally, sterile neutrinos from resonant production can be fully disfavoured as a dominant DM candidate. The results of this paper show that the 21-cm absorption signal is a powerful discriminant of non-cold dark matter, allowing for significant improvements over to the strongest current limits. Confirming the result from EDGES is paramount in this context.
\end{abstract}

\maketitle

\section{Introduction}
While the standard model of cosmology ($\Lambda$CDM) provides an accurate description of the large scale structures, there is still considerable uncertainty at the scales of dwarf galaxies and below. Many alternative dark matter (DM) scenarios predict suppressed perturbations at these scales, resulting in fewer dark matter haloes with generally flatter profiles. Prime examples are thermally produced warm or mixed DM \citep{Bode:2000gq,Boyarsky:2008xj,Schneider:2014rda}, sterile neutrinos \citep{Boyarsky:2009ix,Adhikari:2016bei}, ultra-light axions \citep{Hu:2000ke,Marsh:2013ywa}, or interacting DM models \citep{Boehm:2014vja,Cyr-Racine:2015ihg,Escudero:2018thh}. Apparent tensions between CDM predictions and observations based on gravity-only simulations of dwarf galaxies have further motivated such alternative scenarios \citep{Weinberg:2013aya,Schneider:2016ayw}. However, during the last decade it has become more and more evident that baryonic effects driven by supernova feedback and high-redshift reionisation have the potential to solve most of these tensions [e.g. \citealp{Brooks:2012vi,DiCintio:2013qxa,Sawala:2015cdf}; but see also \citealp{Trujillo-Gomez:2016pix,Schneider:2017vfo}]

Independently of whether alternative dark matter models provide a better match to the data, it is possible to constrain them with astrophysical observations. The currently strongest limits come from the Lyman-$\alpha$ forest constraining the thermal particle mass of warm dark matter (WDM) to $m_{\rm TH}\gtrsim3.5$ keV \citep{Viel:2013apy,Irsic:2017ixq}\footnote{Even more restricting limits from the Lyman-$\alpha$ forest \citep{Baur:2015jsy,Irsic:2017ixq} of the order of $m_{\rm TH}\gtrsim5$ keV are strongly disputed due to uncertainties in the temperature evolution \citep{Garzilli:2015iwa} and radiation effects of the intergalactic medium \citep{Keating:2017lgk}.}. Other constraints on the WDM mass from Milky-Way satellites \citep{Polisensky:2010rw,Lovell:2011rd,Kennedy:2013uta,Schneider:2014rda}, high-redshift galaxies \citep{Menci:2016eui,Corasaniti:2016epp}, or strong gravitational lensing \citep{Birrer:2017rpp, Vegetti:2018dly} are currently around $m_{\rm TH}\sim1.5-3$ keV.

Recently, the {\it Experiment to Detect the Global Epoch of Reionization Signature} (EDGES) reported a strong absorption signal at $\nu\sim 78\pm1$ MHz relative to the cosmic microwave background (CMB) radiation \citep{Bowman:2018yin}. At this frequency, any absorption trough is expected to be induced by the UV light of the first radiative sources, which alter the excitation state of the 21-cm hyperfine transition via the Wouthuysen-Field effect \citep{Wouthuysen:1952aaa,Field:1958aaa}. Assuming standard physics, the amplitude of the signal is bracketed by the CMB and the kinetic gas temperature, and should therefore be of order $200$ mK or below. However, the signal reported by EDGES is more than a factor of two larger, which means that new physics is required to explain its amplitude \citep{Barkana:2018lgd}. Several possibilities have been put forward, such as additional gas cooling via interactions with dark matter \citep[e.g.][]{Munoz:2018pzp,Barkana:2018qrx,Slatyer:2018aqg} or a high-redshift radio source amplifying the CMB radiation \citep{Feng:2018rje} which is, however, likely to be of exotic origin \citep{Sharma:2018agu}.

In the present paper we do not discuss the amplitude of the absorption trough but we solely focus on the timing of the signal. The reported frequency of $\nu\sim 72-85$ MHz translates into a redshift range of $z\sim15.5-19.5$ at which sufficient UV radiation has to be present to induce a signal. Since star formation requires collapse of gas within the potential wells of dark matter haloes, any model with delayed halo formation can be constrained using the global 21-cm signal. This has been shown explicitly in the past for the case of of thermally produced WDM \citep{Barkana:2001gr,Sitwell:2013fpa,Bose:2016hlz,Villanueva-Domingo:2017lae,Lopez-Honorez:2017csg,Escudero:2018thh}. Motivated by the signal from EDGES, we perform a detailed analysis of how the 21-cm signal depends on halo formation and the nature of dark matter. Additionally to WDM, we also discuss ultra-light axion DM, sterile neutrinos, and mixed DM with a cold and a warm/hot component.

The paper is structured as follows. In Sec.~\ref{sec:model} and \ref{sec:massfct} we discuss key aspects of the global 21-cm signal with specific focus on the role of dark matter. In Sec.~\ref{sec:signal} and \ref{sec:constraints} the predicted models are compared to the timing of the signal from EDGES, resulting in constraints on various DM particle models. Throughout the paper, we assume a {\tt Planck} cosmology with $\Omega_{\Lambda}=0.685$, $\Omega_m=0.315$, $\Omega_b=0.049$, $h=0.673$, $n_s=0.965$, and $\sigma_8=0.83$ \citep{Planck:2015xua}.

\section{The model}\label{sec:model}
The differential brightness temperature of the 21-cm signal is given by the difference between the spin temperature of of the gas ($T_s$) and the source temperature form the cosmic microwave background ($T_{\gamma}$), i.e.
\be\label{dTb}
\delta T_b \simeq 27 x_{\rm HI}\left(\frac{\Omega_bh^2}{0.023}\right)\left(\frac{0.15}{\Omega_m h^2}\frac{1+z}{10}\right)^{\frac{1}{2}}\left(1-\frac{T_{\gamma}}{T_s}\right)
\ee
in Milli-Kelvin, where $x_{\rm HI}$ is the neutral gas fraction which is very close to one for all redshifts of interest in the present study \citep[see e.g. Refs.][]{Furlanetto:2006tf,Pritchard:2011xb}.

The spin and kinetic gas temperatures are related via the equation
\be\label{Ts}
\left(1-\frac{T_{\gamma}}{T_s}\right)\simeq\frac{x_{\rm tot}}{1+x_{\rm tot}}\left(1-\frac{T_{\gamma}}{T_{\rm k}}\right),
\ee
with $x_{\rm tot}=x_c+x_{\alpha}$ being the sum of the collisional and radiative coupling parameters (see definition below). The gas temperature evolves according to the relation
\be\label{Tk}
\frac{dT_{\rm k}}{dt}+2HT_{\rm k} = \frac{2}{3 k_{B}n_{\rm tot}}\sum_i\Gamma_i\,,
\ee
where $H=H(z)$ is the Hubble parameter, $n_{\rm tot}$ is the gas density, and $\Gamma_i$ are the different heating and cooling rates \citep[including Compton and X-ray heating, see e.g.][]{Furlanetto:2006tf}.

Based on Eqs.~(\ref{dTb}-\ref{Tk}), we can sumarise the evolution of the observable 21-cm signal. Below $z\sim200$ the gas decouples from the CMB (i.e. the Compton heating becomes inefficient) and $T_k$ cools adiabatically, falling below the CMB temperature at a rate $T_{\gamma}/T_{k}\propto(1+z)$. A first absorption signal is expected at these redshifts, since the collisional coupling coefficient ($x_c>0$) drives $\delta T_b$ towards $T_{\rm k}$. Around $z\sim40$, collisional coupling becomes inefficient ($x_c=0$) and the absorption signal vanishes again. Later on, after the formation of the first stars, the radiative coupling coefficient ($x_{\alpha}$) becomes non-zero, leading to a second absorption feature below $z\sim 30$. This lasts until the X-ray radiative background heats up the gas, transforming the absorption into an emission signal.

\begin{figure*}
\centering{
\includegraphics[width=.245\textwidth,trim={0.05cm 0.8cm 1.2cm 0.2cm}]{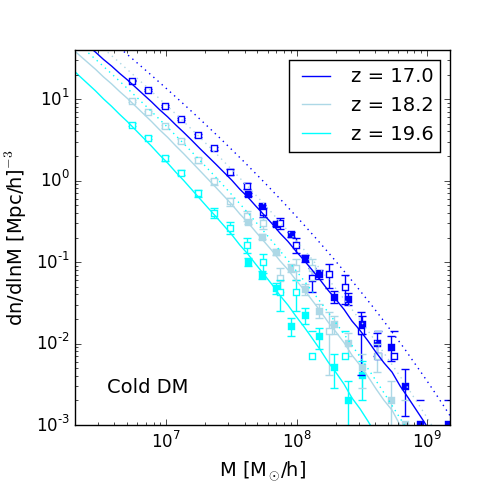}
\includegraphics[width=.245\textwidth,trim={0.05cm 0.8cm 1.2cm 0.2cm}]{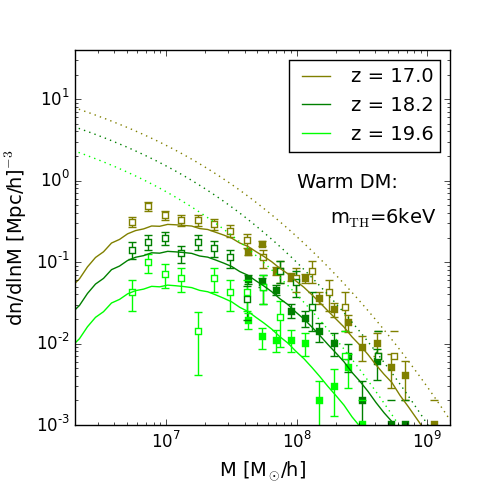}
\includegraphics[width=.245\textwidth,trim={0.05cm 0.8cm 1.2cm 0.2cm}]{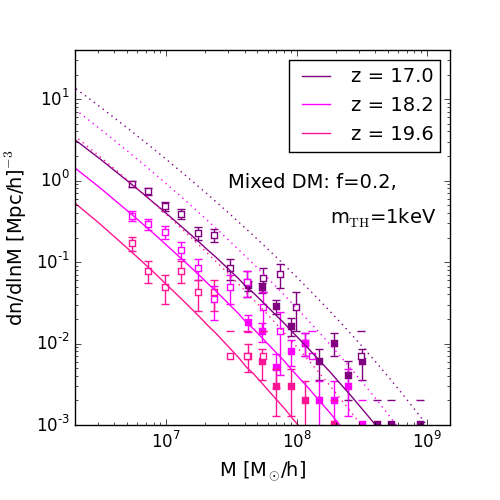}
\includegraphics[width=.245\textwidth,trim={0.05cm 0.8cm 1.2cm 0.2cm}]{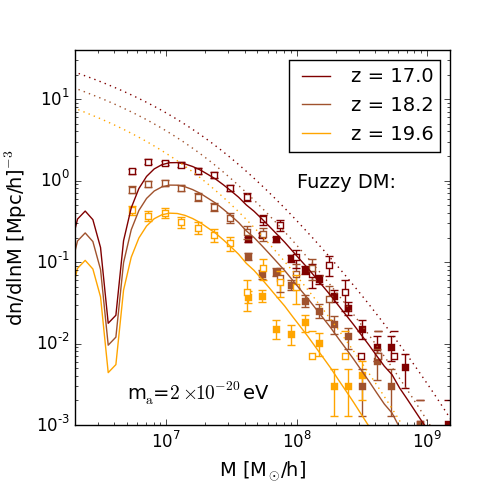}
}
\caption{\small{Halo mass functions at redshifts 17, 18.2 and 19.6 for cold, warm, mixed, and fuzzy (ultra-light axion) dark matter (from left to right). Empty and filled symbols are from N=$1024^3$ simulations with box-length of 8 and 16 Mpc/h, respectively. Error bars correspond to the Poisson uncertainties. Solid and dotted lines show the predictions from the extended Press-Schechter model with sharp-$k$ and tophat filter.}}
\label{fig:massfct}
\end{figure*}

The UV coupling coefficient ($x_{\alpha}$) is given by
\be\label{couplingcoefficient}
x_{\alpha}=1.18\times 10^{11}\frac{J_{\alpha} S_{\alpha}}{(1+z)},
\ee
where $J_{\alpha}$ is the Lyman-$\alpha$ flux (with units  cm$^{-2}$ s$^{-1}$ Hz$^{-1}$ sr$^{-1}$) and $S_{\alpha}$ is a dimensionless factor that accounts for spectral distortions. We assume $S_{\alpha}=1$ which is a conservative choice \citep{Hirata:2005mz}. The Lyman-$\alpha$ flux is given by
\be
J_{\alpha}= \frac{(1+z)^2}{4\pi}\sum_{n=2}^{n_m}f_{n}\int_{z}^{z_{\rm max,n}} dz'\frac{c}{H(z')}\epsilon_{\nu} (z'),
\ee
where the sum is truncated at $n_m=23$ and where $f_n$ represent the recycling fractions \citep[see][]{Barkana:2004vb,Pritchard:2005an}. The integration limits are given by $(1+z_{\rm max})= (1+z)\left[1-(n+1)^{-2}\right]/\left[1-n^{-2}\right]$. The emissivity parameter $\epsilon_{\nu}(z)$ can be modelled via the relation \citep{Barkana:2004vb}
\be\label{emissivity}
\epsilon_{\nu}(z)=\frac{N_{\alpha}}{(\nu_{\rm LL} - \nu_{\alpha})m_b}\dot{\rho}_{*}(z),
\ee
where $\dot{\rho}_{*}$ is the star-formation rate density, $m_b$ the proton mass, and $N_{\alpha}$ the total number of emitted photons per stellar baryon in the range between the Lyman-$\alpha$ and Lyman limit frequencies (i.e. $\nu_{\alpha}$ and $\nu_{\rm LL}$). We assume $N_{\alpha}=9690$ corresponding to the yield of population II stars which is more than two times larger than the one of population III stars \citep{Pritchard:2005an}. The star-formation rate is proportional to the accreted matter, i.e.,
\be\label{sfr}
\dot{\rho}_{*}(z)=f_{*}\bar{\rho}_{b,0}\frac{d}{dt}f_{\rm coll}(z),
\ee
where $f_{*}$ is the fraction of gas transformed into stars, $\bar{\rho}_{b,0}$ the mean baryon density at $z=0$, and $f_{\rm coll}(z)$ the amount of matter in haloes. The latter can be obtained by integrating the halo mass function as follows
\be\label{fcoll}
f_{\rm coll}(z)=\frac{1}{\rho_m}\int_{M_{\rm min}}^{\infty}dM \frac{dn}{d\ln M}\,,
\ee
where $M_{\rm min}$ is the minimum halo mass below which no gas cooling is expected. We will now discuss suitable choices for $M_{\rm min}$ and $f_{*}$ referring to the next section for a model of the halo mass function. 

At very high redshifts, gas can cool via the atomic cooling mechanism in haloes with mass above $M_{\rm min}\sim10^7$ $M_{\odot}$/h. Below this threshold, the halo potentials are not deep enough to allow the gas to be shock-heated above $T_{\rm min}\sim 10^4$ K, making the atomic cooling channel ineffective. Molecular $H_2$ cooling works down to $T_{\rm min}\sim 10^3$ K (corresponding to $M_{\rm min}\sim 3\times 10^5$ $M_{\odot}$/h) but $H_2$ molecules can get easily destroyed by radiation. Recent cosmological radiation-hydrodynamic simulations have shown that molecular cooling could indeed play a crucial role in high-redshift star formation, enabling the build-up of galaxies in haloes below the atomic cooling limit \citep{Wise:2007cf,Wise:2014vwa}. In our analysis, we therefore assume $T_{\rm min}\sim 10^3$ K as lower limit for star formation.

The fraction of gas transformed into stars ($f_{*}$) depends on the details of gas cooling, star formation, and feedback. These processes are still not understood in detail, making $f_{*}$ the largest uncertainty of our analysis. In general, the stellar fraction is expected to depend on halo mass with a peak around $M\sim10^{11}$ M$_{\odot}$/h and a steep decline towards smaller masses due to both feedback and inefficient gas cooling \citep[see e.g. Ref.][for a mass dependent parametrisation of the stellar fraction]{Mirocha:2016aaa}. While abundance matching provides indirect evidence for the decline of $f_{*}$ at redshifts below $z\sim10$, no direct information about the stellar-to-baryon connection is available for higher redshifts \citep[see Refs.][for discussions regarding the very weak constraints on $f_*$ from observations]{Greig:2015qca,Cohen:2016jbh}. 

In this paper we assume a constant value for the stellar-to-baryon fraction with a best-guess value of $f_{*}=0.01$ and a conservative upper limit of $f_{*}=0.03$. This is in agreement radiation-hydrodynamic simulations of high-redshift galaxies in a neutral medium. For example, \citet{Wise:2014vwa} find a stellar-to-baryon ratio consistently below $f_{*}=0.03$ for haloes in the relevant mass range of $10^6-10^9$ M$_{\odot}$/h. This is confirmed by various other simulations, see e.g. Refs.~\citep{Xu:2016aaa,Ma:2017avo,Rosdahl:2018aaa}. The main reason for the low value of $f_*$ is radiation pressure and supernova feedback regulating the formation of stars. No comparable simulations currently exist for the case of non-cold dark matter. However, we do not expect large differences since star formation is driven by astrophysical processes that are unlikely to be strongly affected by DM properties \footnote{Recently, it was argued that star formation could be more efficient in non-cold DM models compared to pure CDM \citep{Lovell:2017eec}. Although the simulations that led to these findings do not account for radiation feedback and lack the resolution to probe the most relevant scales below $10^9$ M$_{\odot}$, they could point towards an interesting effect at the DM cutoff scale. It would be interesting to see if such an effect is visible in more realistic radiation-hydrodynamic simulations covering smaller scales and higher redshifts. Regarding our upper limit on $f_*$, note that it is conservative enough to incorporate the increased star formation found in Ref.~\citep{Lovell:2017eec}.}.

So far we have discussed the emergence of a 21-cm absorption signal induced by the UV light of first stars. We now turn our attention towards the gas heating process which makes the absorption signal disappear again. The gas heating is caused by the X-ray radiation background from starburst galaxies, quasar, and supernova remnants. We adopt a simple recipe for the heating rate $\Gamma_X$ (see Eq.~\ref{Tk}) given by
\be\label{xrayheating}
\Gamma_X(z)= f_{X}f_{\rm heat}c_X\dot{\rho}_{*}(z),
\ee
where $c_X=2.6\times 10^{39}$ erg s$^{-1}$ (M$_{\odot}$yr)$^{-1}$ is a normalisation factor \citep[constrained by observations of the nearby universe, see][]{Mineo:2012aaa}, $f_{\rm heat}$ is the fraction of radiation deposited as heat \citep[obtained as in][]{Shull:1985aaa}, and $f_X$ is an efficiency parameter, absorbing uncertainties related to the redshift evolution. In Ref.~\citep{Mirocha:2014faa} it is shown that such a simple prescription is sufficiently accurate for our analysis. We allow $f_X$ to vary within the limits $0.2\leq f_{X}\leq4$ which produces a signal expected from source galaxies with similar properties than the observed galaxies at $z\sim6-8$ \citep[see Ref.][]{Mirocha:2016aaa}. Also note that a larger value of $f_X$ strongly reduces the overall amplitude of the absorption trough, making it increasingly difficult to reconcile the model with the EDGES signal.

\section{The halo mass function}\label{sec:massfct}
The abundance of haloes as a function of mass and redshift is a crucial ingredient of the model outlined above. For the case of CDM, the halo mass function is well described by the extended Press-Schechter (EPS) method \citep{Press:1973iz,Bond:1990iw,Sheth:2001dp}. However, the standard EPS model fails for non-cold DM models where significant free streaming or particle interactions lead to a suppression of the linear power spectrum. Such models are much more accuratey described by an EPS model with sharp-$k$ filter \citep[][]{Benson:2012su,Schneider:2013ria,Schneider:2014rda,Leo:2018odn}. We follow Refs.~\citep{Schneider:2013ria,Schneider:2014rda} and write
\be\label{massfct}
\frac{dn}{d\ln M}=\frac{1}{12\pi^2}\frac{\bar\rho}{M}\nu f(\nu)\frac{P_{\rm lin}(1/R)}{\delta_c^2 R^3},
\ee
with $f(\nu) = A\sqrt{2\nu/\pi}(1+\nu^{-p})e^{-\nu/2}$, $\nu=(\delta_c/\sigma)^2$, $A=0.322$, $p=0.3$, and $\delta_c=1.686$. The variance is given by
\be\label{variance}
\sigma^2(R,z)=\int\frac{dk^3}{(2\pi)^3}P_{\rm lin}(k)\Theta(1-kR),
\ee
where $P_{\rm lin}(k)$ is the linear power spectrum and $\Theta$ the Heaviside step-function. Finally, the halo mass is connected to the radius via $M=4\pi\bar\rho(cR)^3/3$ with $c=2.5$.

\begin{figure*}
\centering{
\includegraphics[width=.245\textwidth,trim={0.05cm 0.8cm 1.2cm 0.2cm}]{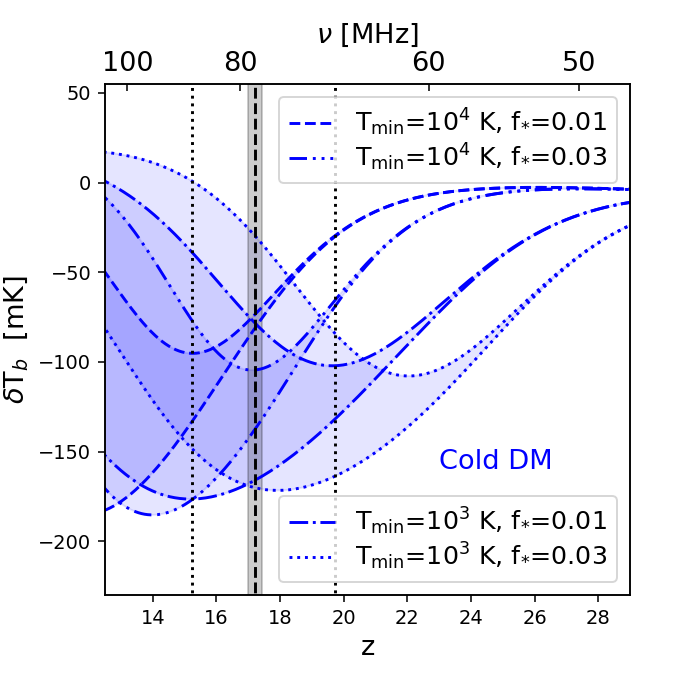}
\includegraphics[width=.245\textwidth,trim={0.05cm 0.8cm 1.2cm 0.2cm}]{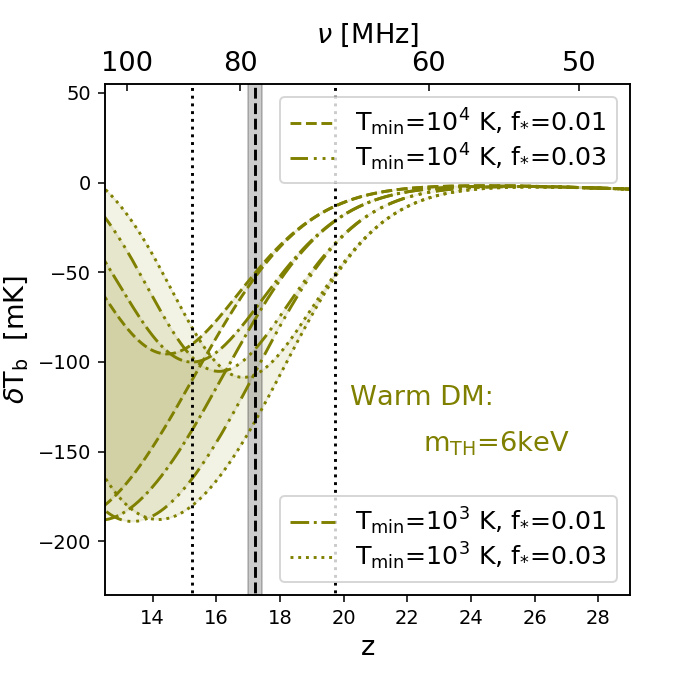}
\includegraphics[width=.245\textwidth,trim={0.05cm 0.8cm 1.2cm 0.2cm}]{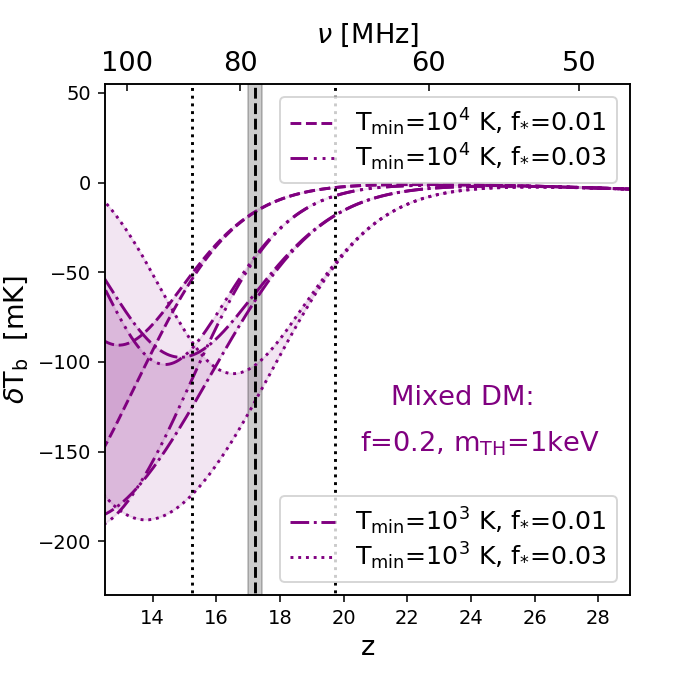}
\includegraphics[width=.245\textwidth,trim={0.05cm 0.8cm 1.2cm 0.2cm}]{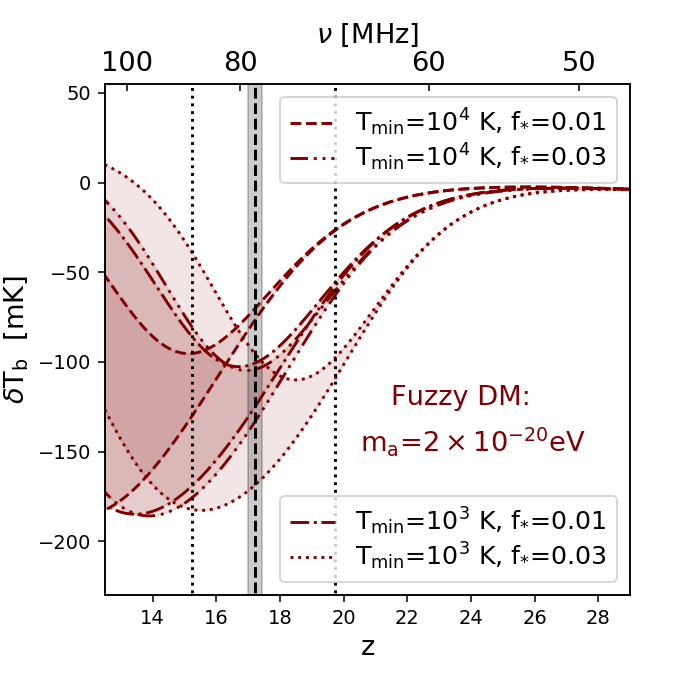}
}
\caption{\small{Absorption signal for cold, warm, mixed, and fuzzy (ultra-light axion) dark matter (from left to right). For each panel, we show four models with different assumptions regarding the stellar-to baryon fraction ($f_*$) and the minimum virial temperature of haloes where gas is able to cool. Solid and dashed lines correspond to cases with best guess and maximum stellar fraction ($f_{*}=0.01,\,0.03$) assuming only atomic cooling ($T_{\rm min}=10^4$ K). Dash-dotted and dotted lines show the same but including molecular cooling (i.e. $T_{\rm min}=10^3$ K). The mean frequency of the EDGES signal \citep{Bowman:2018yin} is shown as vertical dashed line (with the uncertainty delimited by the narrow grey band) while the signal width (where the amplitude is at half of its maximum) is indicated by vertical dotted lines.}}
\label{fig:dTb}
\end{figure*}

The halo mass function of Eq.~(\ref{massfct}) has been shown to provide accurate predictions for generic non-cold DM models at redshift $z\leq 5$ \citep{Schneider:2013ria,Schneider:2014rda,Murgia:2017lwo}. In order to test its applicability for the very high redshifts considered here, we run a suite of $N$-body simulations for cold DM, warm DM (with thermal mass $m_{\rm TH}=6$ keV), mixed DM (with a fraction $f=0.2$ of warm DM of $m_{\rm TH}=1$ keV), and fuzzy DM (i.e. ultra-light axion DM with mass $m_{\rm a}=2\times 10^{-20}$ eV) using the $N$-body code {\tt Pkdgrav3} \citep[][]{Stadsel:2001aaa,Potter:2016ttn}. The initial conditions of the simulations were generated with the {\tt MUSIC} code \citep{Hahn:2011aaa} based on power spectra from {\tt CLASS} \citep{Blas:2011rf,Lesgourgues:2011rh} and {\tt axionCAMB} \citep{Hlozek:2014lca}. Regarding box size and particle numbers, we use $L=8,\,16$ Mpc/h and $N=1024^3$.

Fig.~\ref{fig:massfct} shows the halo mass functions from our simulations (symbols with error bars) together with predictions from the sharp-$k$ as well as the standard tophat EPS mass functions (solid and dotted lines). While the tophat mass function significantly over-predicts the halo abundance for all non-cold DM scenarios, the sharp-$k$ mass function provides a good match to the data. We conclude that Eq.~(\ref{massfct}) can be safely used to predict the clustering of non-cold DM models at the relevant redshifts \footnote{The attentive reader will have noticed the strong oscillatory pattern of the halo mass function for fuzzy DM and wondered if this feature has a physical meaning. At the moment we lack resolution to fully answer this question. However, recent simulations by \citet{Leo:2018odn} suggest that oscillatory features appearing in the power spectrum tend to get erased by nonlinear structure formation. We therefore expect the oscillations to be at least weakened compared to the predictions from the sharp-$k$ model. Concerning the present study, we have checked that this does not affect our results, since they only depend on the integral of the halo mass function.}.

\section{The 21-cm absorption signal}\label{sec:signal}
So far, we have discussed the main steps of the prediction pipeline for the 21-cm absorption signal with an emphasis on how it is affected by the nature of dark matter (DM). In this section we compare the timing of the predicted absorption trough with the observed signal from EDGES. Assuming upper limits on the star-formation rate, this will then allow us to  constrain the DM sector.

The full 21-cm signal as a function of redshift is calculated with the publicly available code  {\tt ARES} \citep[{\it Accelerated Reionization Era Simulations}, see][]{Mirocha:2012aaa,Mirocha:2014faa,Mirocha:2015jra,Mirocha:2016aaa}. We apply a simple setup where the UV emissivity and X-ray heating are computed as in Eqs.~(\ref{emissivity}-\ref{xrayheating}), ignoring both the spectral energy distribution of sources and a potential mass-dependence of the stellar-to-baryon fraction. The halo mass function is calculated separately (following the recipe of Sec.~\ref{sec:massfct}) and used as an input of {\tt ARES}.

While the assumed model is likely too simplistic to capture the details of the global 21-cm spectrum, it is good enough to constrain the {\it timing} of the signal. This means that, by making conservative assumptions on the star-formation rate and the X-ray heating, we can determine the highest possible redshift of the absorption signal.

Fig.~\ref{fig:dTb} shows the differential brightness temperature of the global 21-cm signal for cold, warm, mixed, and fuzzy (ultra-light axion) dark matter (from left to right). Different lines correspond to different assumptions about the minimal gas cooling temperature ($T_{\rm min}$) and the stellar-to-baryon fraction ($f_{*}$). Towards smaller redshifts each line separates in two. The colour-shaded area between the two lines quantifies the uncertainty due to the gas heating processes (assuming $0.2\leq f_X\leq 4$). The model with $T_{\rm min}=10^3$ K and $f_{*}=0.03$ (dotted lines) corresponds to the most extreme case beyond which the assumed star-formation rate is in strong disagreement with radiation hydrodynamic simulations of high-redshift galaxies (see Sec.~\ref{sec:model} for more details). The vertical dashed line shows the average frequency of the EDGES signal (with the error given as grey band). The width of the EDGES signal at half its maximum amplitude is indicated by vertical dotted lines. 

From Fig.~\ref{fig:dTb}, it becomes immediately clear that the timing of the EDGES signal is able to set strong constraints on the nature of dark matter. The very lukewarm DM model shown in the second panel, for example, exhibits absorption troughs consistently shifted to smaller redshifts compared to the EDGES signal. This is in strong contrast to the case of CDM where the uncertainty on the timing is much larger. The reason why the CDM model is more sensitive to astrophysical assumptions compared to WDM can be explained by the fact that reducing $T_{\rm min}$ or increasing $f_*$ only affects the 21-cm signal if there is enough small haloes to start with.

\begin{figure*}
\centering{
\includegraphics[width=.49\textwidth,trim={0.2cm 0.0cm 1.4cm 0.2cm,clip}]{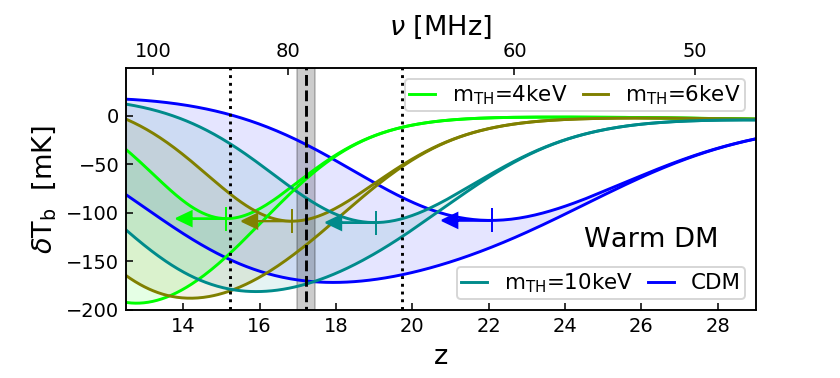}
\includegraphics[width=.49\textwidth,trim={0.2cm 0.0cm 1.4cm 0.2cm,clip}]{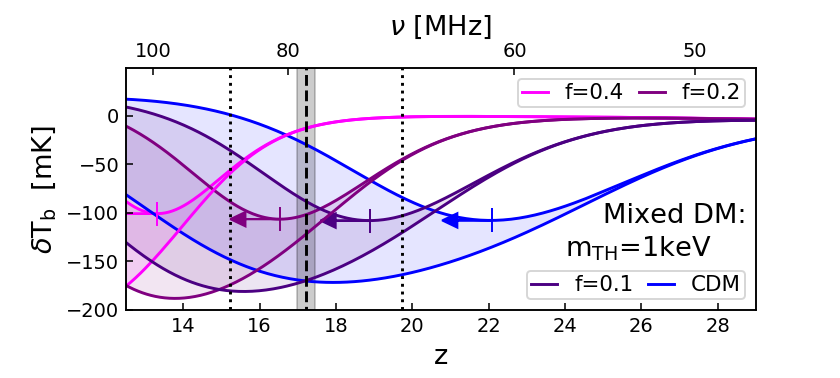}\\
\includegraphics[width=.49\textwidth,trim={0.2cm 0.0cm 1.4cm 0.2cm,clip}]{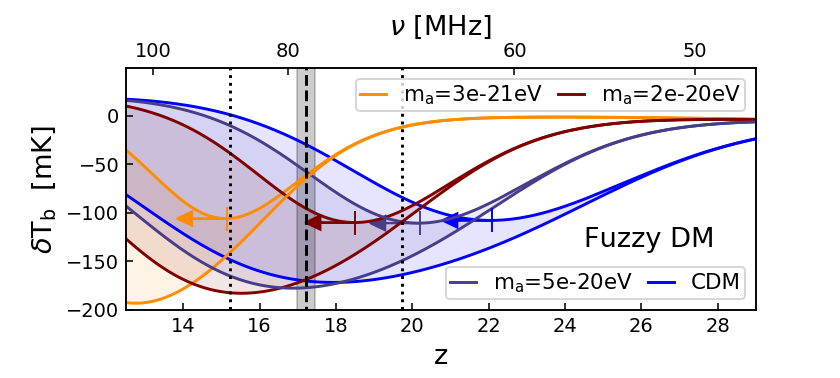}
\includegraphics[width=.49\textwidth,trim={0.2cm 0.0cm 1.4cm 0.2cm,clip}]{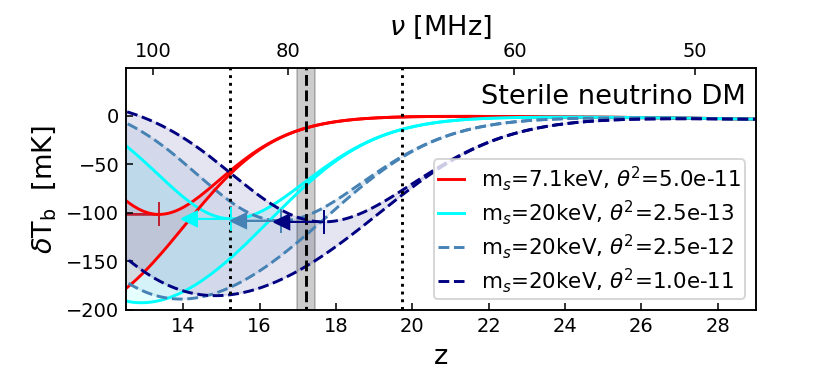}
}
\caption{\small{Absorption signal of various DM models assuming $T_{\rm min}=10^3$ K (including atomic and molecular cooling) and $f_{*}=0.03$ (corresponding to the largest allowed stellar-to-baryon fraction in haloes). From top left to bottom right: thermal warm DM, mixed DM, fuzzy DM, and sterile neutrino DM from resonant production. Coloured arrows illustrate that all absorption signals are allowed to move towards smaller but never towards larger redshifts. Models are excluded if the minimum of their absorption trough is further left than the signal from EDGES (dashed vertical line). Bottom-right panel: Sterile neutrino DM models that are in tension with limits from X-ray observations are shown with dashed lines.}}
\label{fig:dTblimits}
\end{figure*}

\section{Constraining dark matter}\label{sec:constraints}
Our next goal is to derive constraints on the particle properties of DM models. We do this by comparing the frequency (redshift) of the observed and predicted \emph{minimum} of the 21-cm absorption trough. The minimum is a good measure for the timing of the signal because it does not change if additional mechanisms decreasing $T_k$ or increasing $T_{\gamma}$ are assumed (this is only true of course, as long as the mechanism in question has no strong redshift dependence around the minimum). Note that this point is particularly important since the signal measured by EDGES is significantly stronger than expected.

Fig.~\ref{fig:dTblimits} shows the 21-cm absorption troughs of the most extreme model (with $T_{\rm min}=10^3$ K and $f_{*}=0.03$) that represents the limit beyond which the star-formation rate strongly disagrees with results from radiation-hydrodynamic simulations (see discussion in Sec.~\ref{sec:model}). For any realistic model, the minimum of the absorption trough is allowed to shift towards smaller (but never larger) redshifts as indicated by the coloured arrows. Hence, all models of Fig.~\ref{fig:dTblimits} with an absorption minimum at redshifts below $z=17.2$ (corresponding to the signal from EDGES, see vertical dashed line) are excluded. The four panels of Fig.~\ref{fig:dTblimits} show different DM scenarios (with varying model parameters) that we will now discuss in more detail.

The thermal warm DM scenario (top-left panel of Fig.~\ref{fig:dTblimits}) is fully characterised by the particle mass $m_{\rm TH}$. Based on the procedure described above, all models with mass below $m_{\rm TH}=6.1$ keV are in tension with the EDGES signal. This is visible in Fig.~\ref{fig:dTblimits}, where models with smaller $m_{\rm TH}$ have their absorption minima (arrows) to the left of the vertical dashed black line.

Our constraints on WDM are significantly tighter than previous results based on the 21-cm signal \citep{Sitwell:2013fpa,Safarzadeh:2018hhg}. One important reason for this improvement is the sharp-$k$ halo mass function used in our analysis. As shown in Sec.~\ref{sec:massfct}, the sharp-$k$ mass function is in good agreement with simulations of non-cold DM models, while the standard Press-Schechter approach over-predicts the halo abundance towards small masses. Compared to Ref.~\citep{Safarzadeh:2018hhg} who reported a limit of $m_{\rm TH}>3$ keV based on EDGES, we gain further leverage by focusing on the full absorption signal instead of solely the UV coupling coefficient (x$_{\alpha}$). The former allows to directly compare theoretical predictions with observations without relying on the arbitrary choice of a maximum value for $x_{\alpha}$.

The mixed DM scenario (top-right panel of Fig.~\ref{fig:dTblimits}) consists of a composition of both warm/hot and cold dark matter, parametrised by the particle mass of the warm/hot species ($m_{\rm TH}$) and the fraction $f=\Omega_{\rm WDM}/(\Omega_{\rm WDM}+\Omega_{\rm CDM})$. As long as the warm component is sufficiently hot ($m_{\rm TH}\lesssim1$ keV) the 21-cm absorption signal is only affected by the fraction $f$. We obtain a limit of $f\leq0.17$ as indicated in Fig.~\ref{fig:dTblimits}. This means that no more than 17 percent of the DM can be hot without disagreeing with the timing of the EDGES signal. 

The fuzzy DM scenario (bottom-left panel of Fig.~\ref{fig:dTblimits}) consists of an ultra-light boson (i.e. axion-like particle) parametrised by the particle mass $m_{a}$ \citep[][]{Marsh:2015xka,Hui:2016ltb}. Ultra-light axion models are characterised by a large de Broglie wavelength leading to a suppression of the linear power spectrum \citep{Hlozek:2014lca,Sarkar:2015dib} as well as novel features at very nonlinear scales \citep{Schive:2014hza}. For fuzzy DM we obtain a limit of $m_{a}>8\times10^{-21}$ eV. This is confirmed by the models illustrated in Fig.~\ref{fig:dTblimits}, where the orange scenario is ruled out while the brown scenario is still allowed.

Finally, the resonantly produced sterile neutrino DM model  (bottom-right panel of Fig.~\ref{fig:dTblimits}) is characterised by the particle mass ($m_s$) and the mixing angle ($\theta$) with active neutrinos \citep{Shi:1998km,Canetti:2012kh}. Depending on these parameters, a variety of non-thermal particle distribution functions are found \citep[we use the code {\tt sterile-dm} to calculate these distribution functions, see Ref.][]{Venumadhav:2015pla}. The particle mass in the {\it keV} range plus the non-thermal distribution functions result in suppressed linear power spectra with shapes similar to warm or mixed DM \citep{Boyarsky:2008xj}.

With the method developed in this paper, all the remaining parameter space for resonantly produced sterile neutrino DM can be excluded. This is because the parts of the parameter space leading to cold enough power spectra to agree with the timing of the EDGES signal are excluded by X-ray data. In Fig.~\ref{fig:dTblimits} all models excluded by X-ray observations are specifically highlighted with dashed lines. We show several models with varying $\theta$ and fixed particle mass at $m_{s}=20$ keV. This is the largest mass where some of the parameter space is still in agreement with the X-ray limits [see e.g. Fig.~5 in Ref.~\citealp{Schneider:2016uqi}, or Fig.~6 in Ref.~\citealp{Cherry:2017dwu}]. However, Fig.~\ref{fig:dTblimits} shows that all models at this mass range are either excluded by X-ray or in clear tension with EDGES.

Next to the three cases with $m_{s}=20$ keV, we specifically show the model with $m_{s}=7.1$ keV and $\theta^2=5\times10^{-11}$ that naturally reproduces the claimed X-ray line at 3.55 keV \citep[reported by Refs.][]{Bulbul:2014sua,Boyarsky:2014jta}. It is clear from Fig.~\ref{fig:dTblimits} that the 21-cm absorption trough from this model (red line) is in strong tension with the reported timing of the EDGES signal.

\section{Conclusions}\label{sec:conclusions}
In this paper we computed the 21-cm absorption signal of non-cold dark matter (DM) scenarios using a model where the formation of the first stars is linked to the halo accretion. In agreement with previous work \citep{Barkana:2001gr,Sitwell:2013fpa,Safarzadeh:2018hhg}, we find that the absorption signal of non-cold DM models is consistently shifted towards smaller redshifts compared to CDM. This is a natural consequence of the fact that these models predict a delay of halo formation and a reduced abundance of small-scale haloes. Quantitatively, we obtain stronger effects than Refs.~\citep{Sitwell:2013fpa,Safarzadeh:2018hhg} because we rely on the sharp-$k$ halo mass function \citep{Schneider:2013ria,Schneider:2014rda} which includes the turnover of the halo abundance towards small masses and is in better agreement with cosmological simulations than the standard Press-Schechter approach.

Based on results from cosmological radiation-hydrodynamic simulations of high-redshift galaxies within a neutral gas medium \citep{Wise:2007cf,Wise:2014vwa}, we then define conservative limits for the minimum mass of haloes hosting stars ($M_{\rm min}=3.2\times10^5$ M$_{\odot}$/h) and for the maximum stellar-to-baryon fraction ($f_{*}=0.03$). This allows us to put upper limits on the redshift of the 21-cm signal which can then be compared to the redshift of the reported signal from EDGES in order to constrain the DM sector. 

For the thermal warm DM scenario we find a limit of $m_{\rm TH}>6.1$ keV which is significantly stronger than previous constraints from the literature (coming from Lyman-$\alpha$, Milky-Way satellites, high-redshift galaxies, or strong lensing). Similar conclusions can be drawn for the case of fuzzy (ultra-light axion) dark matter were we report a limit of $m_{a}>8\times 10^{-21}$ eV. This is an improvement of more than a factor of two compared to the strongest current constraints from the Lyman-$\alpha$ forest \citep{Irsic:2017yje}, pushing the fuzzy DM scenario to a regime where potential wave-effects are far below the kilo-parsec scale.

For mixed dark matter (consisting of a cold and a warm/hot DM subcomponent), we find that the fraction of warm/hot DM cannot be larger than 17 percent of the total DM abundance. This is independent of the particle mass of the warm/hot component ($m_{\rm TH}$) as long as $m_{\rm TH}\lesssim1$ keV.

For sterile neutrino DM from resonant production we find the entire remaining parameter space (that is still allowed by X-ray observations) to be in tension with the timing of the EDGES signal. This is especially true for the model with $m_s=7.1$ keV and $\theta^2\sim5\times10^{-11}$ which naturally explains the claimed X-ray detection at 3.55 keV \citep[reported by][]{Bulbul:2014sua,Boyarsky:2014jta}.  Note that the above conclusions do not automatically apply to sterile neutrinos from other production mechanisms, most notably scalar decay production \citep{Petraki:2007gq,Merle:2015oja}.

There are several simplifying assumptions going into the analysis of this paper, the most important one being the maximum stellar-to-baryon fraction ($f_*$). We used a value of $f_*=0.03$ which is a factor of $\sim5$ larger than the predictions from cosmological radiative hydrodynamic simulations of Refs.~\citep{Wise:2014vwa,Xu:2016aaa}. Note that the formation of stars is suppressed by feedback from radiation pressure, which is a self-regulating process that does not require fine-tuning. It is very unlikely that the results of these simulations could be changed dramatically without assuming currently unknown sources of UV radiation. Since the stellar-to-baryon fraction is dominated by astrophysical processes, it is furthermore not expected to change significantly for different DM scenarios. We therefore consider our limit on $f_*$ to be conservative.

For those unconvinced by these arguments, we would like to stress that meaningful dark matter constraints can be obtained even for the extreme (and completely unrealistic) assumption of $f_{*}=1$. In this case the limits on warm and fuzzy DM become $m_{\rm TH}>2.8$ keV and $m_{a}>1.2\times 10^{-21}$ eV, respectively. This is still comparable to limits from the Lyman-$\alpha$ forest.

The results obtained in this paper further highlight the potential of the global 21-cm signal as an indirect dark matter probe. The reported absorption trough from EDGES does not only point towards additional mechanisms to either cool down the gas temperature \citep{Barkana:2018qrx,Slatyer:2018aqg} or heat up the radio background \citep{Fraser:2018acy,Pospelov:2018kdh,Dowell:2018mdb}, it also puts strong pressure on any DM model that is characterised by a suppressed power spectrum. An independent confirmation of the signal from EDGES will therefore consist of an important step towards a better understanding of the DM sector.

\begin{acknowledgments}
I want to thank Jordan Mirocha for his help with the {\tt ARES} code. This work is supported by the Swiss National Science Foundation (PZ00P2\_161363).
\end{acknowledgments}

\bibliography{ASbib}

\end{document}